\documentclass[sigconf]{acmart}

\usepackage{booktabs} 
\usepackage{algorithm}
\usepackage[noend]{algpseudocode}
\usepackage{caption}
\usepackage[margin=20pt]{subcaption}
\usepackage{hyperref}
\usepackage{enumerate}
\usepackage{mymacros}
\usepackage{balance} 
\usepackage{paralist}
\usepackage{dsfont}
\usepackage{rotating}
\usepackage{multirow}
\usepackage{dblfloatfix}
\usepackage[moderate,mathdisplays=normal]{savetrees}

\usepackage[font={small},skip=0.5pt]{caption}
\usepackage{setspace}

\usepackage{graphicx}
\usepackage[normalem]{ulem}
\useunder{\uline}{\ul}{}

\usepackage{amsthm}
\theoremstyle{definition}

\copyrightyear{2023}
\acmYear{2023}
\setcopyright{rightsretained}
\acmConference[CIKM '23]{Proceedings of the 32nd ACM International
Conference on Information and Knowledge Management}{October 21--25,
2023}{Birmingham, United Kingdom}
\acmBooktitle{Proceedings of the 32nd ACM International Conference on
Information and Knowledge Management (CIKM '23), October 21--25, 2023,
Birmingham, United Kingdom}
\acmDOI{10.1145/3583780.3614779}
\acmISBN{979-8-4007-0124-5/23/10}

\begin{document}
\title{All about Sample-Size Calculations for A/B Testing:\\ Novel Extensions \& Practical Guide}
   
\author{Jing Zhou}
\affiliation{
  \institution{Apple}
  \city{}
  \country{}
}
\email{jing_zhou6@apple.com}

\author{Jiannan Lu}
\affiliation{
  \institution{Apple}
  \city{}
  \country{}
}
\email{jiannan_lu@apple.com}

\author{Anas Shallah}
\affiliation{
  \institution{Apple}
  \city{}
  \country{}
}
\email{a_shallah@apple.com}

\begin{abstract}
While there exists a large amount of literature on the general challenges and best practices for trustworthy online A/B testing, there are limited studies on sample size estimation, which plays a crucial role in trustworthy and efficient A/B testing that ensures the resulting inference has a sufficient power and type I error control. For example, when sample size is under-estimated, the statistical inference, even with the correct analysis methods, will not be able to detect the true significant improvement leading to misinformed and costly decisions. This paper addresses this fundamental gap by developing new sample size calculation methods for correlated data, as well as absolute vs. relative treatment effects, both ubiquitous in online experiments. Additionally, we address a practical question of the minimal observed difference that will be statistically significant and how it relates to average treatment effect and sample size calculation. All proposed methods are accompanied by mathematical proofs, illustrative examples, and simulations. We end by sharing some best practices on various practical topics on sample size calculation and experimental design. 
\end{abstract}

\ccsdesc[500]{Mathematics of computing~Hypothesis testing and confidence interval computation}

\keywords{A/B testing, online controlled experiments, sample size calculation, correlated data, absolute lift, relative lift, minimally observed difference}

\settopmatter{printfolios=true}
\maketitle

\section{Introduction}
\label{sec:intro}

The past two decades have witnessed A/B testing (online controlled experimentation) \cite{kohavi2007, kohavi2020trustworthy} become the go-to methodology in the information technology industry to evaluate new software products and features \cite{bakshy2013uncertainty, Hohnhold2015, gui2015network, xu2015infrastructure, dmitriev2016measuring, lee2018winner, garcia2018understanding, ju2019sequential, liu2021trustworthy, koning2022experimentation}. A/B testing is a practice of randomly allocating user traffic to either control (``A'') or treatment (``B''), and comparing differences (in e.g., engagement or satisfaction). It quickly gained popularity among the data mining community due to its unique and salient features. First and foremost, like randomized clinical trials, A/B testing is arguably the gold standard for drawing causal conclusions \cite{rubin2008objective}. Second, compared to traditional off-line evaluation \cite{thomas2022crowd}, A/B testing has the potential to accumulate large-scale traffic in real-time. Third, A/B testing by design empowers authenticity by ``listening to your customers'' as pointed out in \cite{kohavi2007practical}.

As companies pursue ``more, better and faster'' A/B tests \cite{googlesurvey}, it is crucial that individual tests are adequately powered, so that actionable insights can be properly extracted and measured. 
Interestingly, while there is a growing literature on increasing sensitivity \cite{deng2013cuped, drutsa2015future, xie2016improving, liou2020variance}, to our best knowledge, there appears to be a lack of in-depth studies on power analysis, also known as sample size calculation. In particular, it is unclear how to calculate sample size for correlated observations (where the unit of analysis is more granular than the unit of randomization) and percentage treatment effect (``relative lift''), both ubiquitous in A/B testing \citep{deng2017trustworthy, deng2018}. In this paper, we aim to fill these gaps, and provide A/B researchers and practitioners with a self-contained sample calculation guide, along with providing solutions to several practical topics related to sample size calculation in experimental design. 

The remainder of the paper is organized as follows. Section \ref{sec:exist} gives the background on the standard sample size calculation for A/B testing where the data are assumed to be independent and separately, the existing analysis approaches when data are correlated. In Section 3, we introduce a novel sample size calculation method that can deal with correlation, which is particularly useful for online collected data.  Section 4 extends the sample size formula when the hypothesis of interest is relative lift rather than absolute lift, and discusses its difference and impact compared to absolute lift. The concept of minimal observed difference (MOD) is discussed in Section 5. All of Sections 3-5 are provided with mathematical derivations, illustrative examples, and simulations. Section 6 touches on a number of practical problems that often occur in A/B experimental design and sample size calculation. Section 7 provides concluding remarks.

\section{Existing Work}
\label{sec:exist}

\subsection{Sample Size for Independent Data}
\label{sec:iid}
Let $X_1, \ldots ,X_n$ be i.i.d. control observations and $Y_1, \ldots ,Y_n$ be i.i.d. treatment observations. Let $\mu_x = \e(X)$ and $\mu_y = \e(Y)$ denote the unknown population means for control and treatment, respectively. 

For continuous outcomes, consider a simple two-sample t-test \citep{student1908probable} to test the null hypothesis $H_0: \mu_x=\mu_y$ versus the alternative hypothesis $H_1: \mu_x \neq \mu_y$. For fixed average treatment effect (ATE) $\delta = \mu_y - \mu_x$, and assuming equal-sized arms, the required sample size for each arm can be estimated by \cite{van2002statistical}:
\begin{equation}
n = 2 \sigma^2 \cdot (z_{1-\alpha/2} + z_{1-\beta})^2 / \delta^2,
\label{iidSS}
\end{equation}
where $\sigma$ is standard deviation of both treatment and control observations, $\alpha$ and $\beta$ are permissible type I/II error rates respectively, $z_k$ is $k^{th}$ percentile of the standard normal distribution, and $\delta$ is ATE, also known as MDE (minimal detectable effect). 

It is common to presume the same sample size for treatment and control, given the fact that this guarantees the total sample size is minimized \cite{kohavi2022b}. 
Setting type I error of $\alpha =0.05$ and power of $1-\beta=0.8$ leads to $n \approx 16 \sigma^2 / \delta^2$, a widely adopted rule of thumb in both academia and industry \cite{kohavi2007, kohavi2009controlled}.  Here a 2-sided test is often the hypothesis of interest, although modifying to a 1-sided test is straightforward by replacing $z_{1-\alpha / 2}$ with $z_{1-\alpha}$. 
 
When the outcome is binary, we have the analogous two-sample proportion test with $H_0: p_x=p_y$ vs. $H_1: p_x \neq p_y$ with $\delta = p_y-p_x$.  The corresponding sample size per arm, assuming equal sample size between treatment and control, is \cite{joseph1973statistical}:
\begin{equation}
n = 2 p_\mathrm{pool} (1-p_\mathrm{pool}) \cdot (z_{1-\alpha/2} + z_{1-\beta})^2 / \delta^2,
\label{iidSS_bin}
\end{equation}
where $p_{\textrm{pool}} = (p_x + p_y) / 2$. Notice that the difference between \eqref{iidSS} and \eqref{iidSS_bin} is only how standard deviation is computed. We refer to \eqref{iidSS} and \eqref{iidSS_bin} as the standard sample size calculation hereafter.

\subsection{Analysis Methods for Correlated Data}
\label{sec:deltaAM}

Correlated data are ubiquitous in online controlled experiments, occurring when the unit of randomization (e.g., user or device) is less granular than the unit of analysis (e.g., session) \cite{deng2017trustworthy, liu2023measuring}. For example, consider an experiment where session conversion rate is the primary metric, and the experiment is randomized on users, and each user can have multiple sessions (i.e. multiple rows of records). In this case, sessions do not meet the independence assumption due to the inherent correlation in the data. 
This type of experiment is also referred to as cluster randomization in the statistics and econometrics literature \cite{klar2001, cameron2015practitioner, su2021model}, where ``cluster'' is equivalent to the randomization unit.

Common techniques to analyze correlated data include generalized estimating equations (GEE) \cite{zeger1988, diggle2002}, mixed effects models \cite{galecki2013lme}, or the Delta method \cite{ver2012invented}. Among these, the Delta method has become especially popular in A/B testing, because of its computation efficiency, as described in \cite{deng2018}. 

The Delta method is used as follows. Consider there are $k$ users randomized in a treatment arm, where user $i$ contains $N_i$ observations $X_{ij}$ $(i = 1, ..., k; j = 1, ..., N_i).$ Then the corresponding average metric is $\bar X = \sum_{i,j} X_{ij} \big / \sum_i N_{i}.$ 
It is usually reasonable to assume that different users are independent, but the observations within each user are likely correlated. Hence, ignoring these correlations within each user and treating them as independent would lead to incorrect inferences, more specifically, under-estimation of the variance of $\bar X.$ To address this issue,  \cite{deng2018} first re-wrote $\bar{X}$ as a function of two intermediate i.i.d variables: $S_i$ and $N_i$, and divided the number of users k in both the denominator and the numerator:
\begin{equation}
\label{Xbar}
\bar X = \frac{\sum_{i,j} X_{ij}}{\sum_i N_i} = \frac{\sum_i S_i/ k}{\sum_iN_i / k} = \frac{\bar S}{\bar N},
\end{equation}
where $S_i = \sum_j X_{ij}$ denotes the sum of all observations for $i^{th}$ user. Then second, utilized the bi-variate Delta method to obtain: 
\begin{equation}
    \var(\frac{\bar{S}}{\bar{N}}) \approx \frac{1}{k \mu_N^2} \left(\sigma_S^2 -2\frac{\mu_S}{\mu_N}\sigma_{SN} +\frac{\mu_S^2}{\mu_N^2}\sigma_N^2\right),
    \label{deltaXbar}
\end{equation}
where $\mu_S = \e(S_i)$, $\mu_N = \e(N_i)$, $\sigma^2_S = \var(S_i)$, $\sigma^2_N = \var(N_i)$ and $\sigma_{SN} = \cov(S_i, N_i)$. In other words, the variance of $\bar X$ can be characterized by the bi-variate distribution of $(S_i, N_i).$

\section{Sample Size for Correlated Data}
\label{sec:deltaSS}
\subsection{Motivation}
It is well-understood that for clustered randomized experiments, ignoring intra-class correlation would lead to more false positives, and the Delta method, as illustrated in Section \ref{sec:deltaAM}, is a powerful tool to address this issue. However, when it comes to sample size calculation, the standard sizing formulas \eqref{iidSS} and \eqref{iidSS_bin} (which assume i.i.d.) are seemingly still used universally, resulting in under-powered experiment designs. Both of the aforementioned issues are due to under-estimation of the variance. There is literature in medical research field for sample size calculations for correlated data using GEE approaches \cite{liu1997,shih1997,pan2001}, but they are rather hard to implement, and require a pre-specified correlation structure to begin with. 

This paper aims to draw awareness to this mismatch between the design and analysis of clustered online experiments and to provide a solution for it. It turns out that, curiously but somewhat expectedly, we can fill this gap by none other than the Delta method itself. In the remainder of this section, we show how using the Delta method consistently for both sample size calculation and test analysis ensures a trustworthy experimental design that meets the desired statistical power and type I error rate.

\subsection{Main Result}
\label{sec:mainResult}
To properly calculate sample size from correlated data, we can re-write \eqref{iidSS} as
$(\sigma^2/n)^{-1} = 2 (z_{1 - \alpha / 2} + z_{1 - \beta})^2 / \delta^2$,
then replace $\sigma^2/n$ with the non i.i.d. version of  $\var(\bar X)$ in \eqref{deltaXbar}, and solve for $k$, resulting in the required number of users:
\begin{equation}
\label{deltaSS}
k = 2 h \cdot (z_{1 - \alpha / 2} + z_{1 - \beta})^2 / \delta^2,
\end{equation}
where
\begin{equation}
\label{h}
h =\frac{1}{\mu_N^2} \left(\sigma_S^2 -2\frac{\mu_S}{\mu_N}\sigma_{SN} +\frac{\mu_S^2}{\mu_N^2}\sigma_N^2\right).
\end{equation}
When designing the experiment, all the required components in $h$ can be estimated easily from historical data by aggregating data to the user-level. Since the Delta method relies on the central limit theorem \cite{van2000clt}, its variance estimation works universally for both continuous and binary outcomes. Therefore, only one universal sample size formula \eqref{deltaSS} is needed, unlike the standard sample size formula, which varies for the two types of outcomes. 

The novelty of the new sample size formula \eqref{deltaSS} is recognizing that the Delta method takes into account the clustering structure and estimates $\var(\bar X)$ as a function of $k$ (i.e. randomization unit). The standard formula \eqref{iidSS} instead focuses on $n$ (i.e. analysis unit) and requires $\sigma^2$ which is impossible to obtain due to the correlation. The new formula provides an alternative way to conveniently compute the number of randomization units needed to power the experiment. Here we use “user” as a typical example of randomization unit, which can be replaced with any granularity level that can be safely assumed independent.

A key distinction between sample size calculations for correlated vs independent data is, $h$ being at the center of \eqref{deltaSS}, while $\sigma^2$ is at the center of \eqref{iidSS}. This difference leads to two subtle but important questions regarding the scale and stability of $h$.

We start with a practical question. Being the variance of i.i.d. observations, $\sigma^2$ is not related to the duration (a.k.a. ``scale'') of an experiment, as long as our estimate of $\sigma^2$ from historical data (be it a week, month or quarter) accurately reflects the new experiment. For correlated data, however, since $h$ relies on $N_i$ (number of sessions for each user), it is not ``scale free'', which is also true for the aforementioned GEE approaches. Consequently, we need to pre-set the duration for the new experiment, estimate the elements of $h$ in \eqref{h} accordingly using the same duration from the historical data, and calculate $k$ by \eqref{deltaSS}. If the resulting \% daily traffic recommendation becomes infeasible, we can extend the duration and go through the same process again. We also verified this non-scale-free property through simulations in Section \ref{sec:simulation}.

Next a more theoretical question, the standard i.i.d. sample size calculation implicitly assumes constant variance $\sigma^2$ between treatment and control\footnote{This assumption is where the ``2'' term in \eqref{iidSS} and \eqref{iidSS_bin} comes from.}. However, for correlated data, $h$ not only relies on $N_i$ and $S_i = \sum_{j = 1}^{N_i}X_{ij}$, but also various non-linear terms in \eqref{h}. Therefore, instead of simply assuming constant $h$ between treatment and control, we mathematically prove equivalence or quantify the difference, under clearly specified assumptions. 
\begin{theorem}
\label{thm:equiv}
Let $\mu_i$ and $\sigma_i$ be the mean and standard deviation of the observations from user $i:$
$$
X_{ij} \mid \mu_i, \sigma_i \stackrel{\mathrm{iid}}{\sim} [\mu_i, \sigma_i]
\quad
(j = 1, \ldots, N_i).
$$
Under the assumption that for $i = 1, \ldots, k$, the treatment has no effect on $N_i$ and $\sigma_i$, and a constant effect $\delta$ on $\mu_i$, $h$ in \eqref{h} is a constant between treatment and control. 
\end{theorem}

We briefly discuss the assumptions made:
first, the constant effect on $\mu_i$ is the imperative for sample size calculation, as heterogeneity (without specific modeling assumptions) makes the problem generally intractable.
Second, the ``stable denominator'' assumption (i.e., no effect on $N_i$) is reasonable under incremental treatment effects, commonly expected in practice\footnote{In the rare case of a large user behaviour change, we recommend evaluating the denominator $N_i$ by number of sessions per user as an A/B metric, and should a statistically significant difference occur, either calculating $h$ separately for treatment and control to acknowledge the power impact post-experiment, if any, or alternatively using converted sessions per user as the primary metric can be more suitable.} \cite{deng2017trustworthy, azevedo2020b}.
Finally, although the assumption of no effect on $\sigma_i$ is generally reasonable for continuous outcomes, it does not hold for binary outcomes, because $\sigma_i^2 = p_i (1 - p_i).$ Fortunately, we can exactly quantify the difference of $h$ between treatment and control.

\begin{theorem}
\label{thm:bin}
Let $p_i$ be the mean of binary observations for user $i:$
$$
X_{ij} \mid p_i \stackrel{\mathrm{iid}}{\sim} \mathrm{Bernoulli}(p_i)
\quad
(j = 1, \ldots, N_i).
$$
Under the assumption that for $i = 1, \ldots, k$, the treatment has no effect on $N_i,$ and a constant effect $\delta$ on $p_i$, the difference of $h$ between treatment and control is 
$$
\delta_h = \delta \frac{(1 - \delta) \hat \mu_N - 2 \hat \mu_S}{\hat \mu_N^2}. 
$$
\end{theorem}

In practice, $\delta$ is typically small and $\mu_S$ is bounded by $\mu_N.$ Consequently, $\delta_h \approx 0.$ This is also validated through extensive simulations in Section \ref{sec:simulation}.

We prove Theorems \ref{thm:equiv} and \ref{thm:bin} in the Appendix.

\subsection{Illustrating Example}
\label{sec:deltaSSeg}
This subsection will demonstrate through an example how the sample size and daily traffic allocation can be computed. Consider a user-randomized experiment with session conversion rate as the primary metric, the steps are:
\begin{enumerate}
    \item[i.] Determine a desired duration for the new experimentation. We will use two weeks as an example; 
    \item[ii.] Collect session-level data from existing traffic from the most recent two weeks containing columns of (a) user identifier, (b) session identifier is optional, and (c) primary metric recorded at each session (e.g. binary indicator of whether a session is converted): $X_{ij}$, where $i$ is user index, $j= 1, ..., N_i$ is session index within $i^{th}$ user;
    \item[iii.] Aggregate the data to user-level: (a) count number of rows per user, $N_i$; and (b) sum up metric values for all the sessions for each user: $S_i = \sum_{j} X_{ij}$, leading to something similar to Table \ref{t:deltaSSeg} as a toy example, although there will be thousands of users, if not more in reality;
    \item[iv.] Estimate each component in $h$ using column $N_i$ and $S_i$. Using Table \ref{t:deltaSSeg}, for example:  
    \begin{itemize}
        \item $\hat{\mu}_S = 2.6$: average number of converted sessions per user; 
        \item $\hat{\mu}_N = 4.2$: average number of sessions per user; 
        \item $\hat{\sigma}^2_S = 10.3$: sample variance of number of converted sessions across users; 
        \item $\hat{\sigma}^2_N = 12.7$: sample variance of number of sessions across users;
        \item $\hat{\sigma}_{SN} = 10.1$: sample covariance between number of converted sessions and number of sessions, across users. 
    \end{itemize}
    With the above we have $h = 0.151$;
    \item[v.] Assume an average treatment effect of 5\% increase, with 80\% power and 2-sided 5\% type I error, using \eqref{deltaSS} yields the required number of users $k = 949$. 
    \item[vi.] From the user-level aggregated data from step iii, the number of unique of users can be also computed. For example, if there are 9,500 rows in Table \ref{t:deltaSSeg}, this indicates that 10\% traffic needs to be allocated per arm for the new experiment and run for two weeks. However, if the most recent two weeks only have 900 unique users, say, then a longer duration should be planned and the process restarts from step i until a reasonable traffic percentage is achieved. Overtime, a more reasonable first-guess duration can be learned from practicing this approach for any metrics and business area of interest, and re-computation can be avoided.
\end{enumerate}

\begin{table}[!th]
\caption{Toy example of aggregated user-level data}
\begin{tabular}{ccc}
\hline
User ID  &  $N_i$ (\#Sessions)  &  $S_i$ (\#Converted Sessions) \\
\hline
A&	1&	1\\
B&	3&	3\\
C&	5&	1\\
D&	2&	0\\
E&	10&	8\\
\hline
\end{tabular}
\label{t:deltaSSeg}
\end{table}

As a comparison, if we ignore the correlation and use the standard sample size calculation (\ref{iidSS_bin}) instead, from Table \ref{t:deltaSSeg}, we have $p_x = \sum_i S_i / \sum_i N_i = 0.619$, with the same 5\% absolute lift in $H_1$ leading to $p_\mathrm{pool} = (p_x + p_y) / 2 = 0.644$, and by \eqref{iidSS_bin} we obtain the required number of sessions $n = 1,440$. However, we would roughly collect $k \cdot \mu_N = 949 \cdot 4.2 = 3,986$ sessions using the Delta method, which is much higher than $1,440$ sessions assuming i.i.d., a clearly under-powered experiment.  

We have started the practice of using the proposed sizing method on real large-scale online data for designing A/B tests. Since then, we also found similar patterns as in the illustrating example. For instance, a 1-week experiment on 10\% daily traffic from a standard i.i.d. sizing design \eqref{iidSS_bin} instead required 4 weeks when using the proposed method \eqref{deltaSS}, suggesting the existence of a non-trivial correlation which can significantly impact sample size calculation.

\subsection{Simulation Studies}
\label{sec:simulation}
We have illustrated how the new sample size estimation works in action. This section will verify through simulations that the proposed method robustly ensures power and type I error as claimed and we will also compare it to the performance if using the standard sample size estimation.  We still use a user-randomized experiment with session conversion rate as the motivating example. 

Before using the proposed sample size method, we need to simulate correlated data as if they are the historical data before an experiment. We start with true rate for control as $p_x = 0.6$. Using a clustering structure similar to a random effect model for correlated data, each user $i$ follows its own binary distribution with mean $p_i$ where the correlation is implicitly introduced, and each treatment arm has its common mean $p_x$ centered from all $p_i$ within the arm. Thus, assuming $K = 5,000$ users are used from the historical data, we
\begin{enumerate}
    \item[(a)] generate number of sessions for each user $i$, $N_i \sim \mathrm{Poisson}(\lambda=5)$,
    \item[(b)] under each $i^{th}$ user, generate $p_i \sim N(\mu = p_x = 0.6, \sigma = 0.175)$ truncated at two standard deviation units from the mean to ensure $p_i$ bounded between 0 and 1,
    \item[(c)] for each session $j$ under $i^{th}$ user, let $X_{ij} \sim \mathrm{Bernoulli}(p_i)$ for $j = 1, ..., N_i$ and hence number of converted sessions for each user $i$ will be $S_i = \sum_j{X_{ij}}$. 
\end{enumerate}
After the data are generated, compute all the necessary components in \eqref{h} resulting in $h = 0.071$. With 80\% desired power, 0.05 type I error and assuming a true treatment difference of $\delta = p_y - p_x = 0.05$, we obtain $k = 448$ users required per arm from (\ref{deltaSS}).

Now using $k = 448$ per arm, based on the same aforementioned data generating mechanism, we can simulate control and treatment arm data separately, with the control using $\mu = p_x$ in step (b) and the treatment using $\mu = p_y$ instead. This constitutes one data set, and we repeat this process and independently generate 10,000 new data sets under the null and the alternative respectively to assess the performance of type I error and power. The true treatment difference is 0 under the null with $p_x = p_y = 0.6$ for both arms, and is 0.05 under the alternative with $p_x = 0.6$, and $p_y = 0.65$.  

For each generated data set, we can compute the point and variance estimate of the metric using the Delta method shown in \eqref{Xbar} and \eqref{deltaXbar} in Section \ref{sec:deltaAM} for both control and treatment arms separately. In the end, the significance for each data set can be determined by checking if the corresponding t-statistic larger than the critical value, i.e. 
$\left|\frac{\bar{Y} -\bar{X} }{\sqrt{\var(\bar{Y}) + \var(\bar{X}})}\right| > z_{1-\alpha/2}$.
Aggregating across 10,000 data sets under $H_0$ and $H_1$ gives satisfactory type I error rate of 4.9\% and power of 80.9\%, respectively.

In comparison, if we compute the sample size using i.i.d. method \eqref{iidSS_bin} instead, $n=1,472$ sessions are required, which is obviously smaller than our proposed method suggesting 2,246 sessions (i.e. $k\cdot \mu_N$). In order to assess the corresponding power and type I error, we considered two scenarios to meet 1,472 sessions: (i) maintain number of sessions per user ($N_i$), but number of users $k$ is reduced from the proposed method; and (ii) maintain $k = 448$ users from the proposed method, but reduce $N_i$ proportionally. The corresponding power drops down to 75.8\% in scenario (i) and 68.1\% in scenario (ii), and scenario (i) additionally has inflated the type I error to 11.4\%, while scenario (ii) is reasonable with 5.1\%.  We broadened the simulation cases as shown in Table \ref{t:simu_setup} including varying average number of sessions per user in step (a) from $\lambda=5$ to 20,  true mean difference from 0.05 to 0.02 and 0.2 under the alternative, truncated normal for $p_i$ to uniform with the same range. Then we follow the same procedure as above and report the results in Table \ref{t:simu_delta} for the proposed method and Table \ref{t:simu_iid} for the standard method.

\begin{table}[htbp]
  \centering
  \caption{Simulation Setup}
    \begin{tabular}{lcccc}
    \toprule
       Case   & True $p_x$ & True $\delta$ in $H_1$ & $\lambda$ & $p_i$ distribution (range) \\
    \midrule
    I & 0.6   & 0.05  & 5     & Trunc N(p $\pm$ 0.35) \\
    II & 0.6   & 0.05  & 20    & Trunc N(p $\pm$ 0.35) \\
    III & 0.8   & 0.02  & 5     & Trunc N(p $\pm$ 0.18) \\
    IV & 0.5   & 0.2   & 5     & Trunc N(p $\pm$ 0.3) \\
    V & 0.6   & 0.05  & 5     & Unif(p $\pm$ 0.35) \\
    \bottomrule
    \end{tabular}
  \label{t:simu_setup}
\end{table}
Throughout all the simulations cases I-V, under the required sample size from our proposed method, Table \ref{t:simu_delta} shows that the power and type I error are well maintained. 
Column $k \cdot \mu_N$ in Table \ref{t:simu_delta} and column $n$ in Table \ref{t:simu_iid} can be used to directly compare the difference for sessions required between our approach and the standard approach, which immediately suggests that the standard method under-estimated the sample size. In Table \ref{t:simu_iid}, the finding of scenario (i) confirms that our proposed method is undoubtedly more reliable, and the standard sample size method suffers both power and type I error. Scenario (ii), on the other hand, verifies the practical question indicated in Section \ref{sec:mainResult} that not only the number of required users $k$ is important, but also the duration which determines the average number of sessions per user. When the duration for the experiment is shorter than that used for sample size calculation, power suffers even more than scenario (i). 
From case II, it is not surprising to see the power and type I error impact is more severe when there are more sessions per user compared to case I, which indicates more correlation inherited in the data is compromised by using the i.i.d. sample size method.

\begin{table}[htbp]
  \centering
  \caption{Sample Size \& Performance Using Proposed Method}
    \begin{tabular}{lcccc}
    \toprule
    Case   & $k$   & $k\cdot \mu_N$ & Type I error  & Power \\
    \midrule
    I & 448   & 2,246  & 0.0489 & 0.809 \\
    II & 228   & 4,546  & 0.0514 & 0.816 \\
    III & 1,520  & 7,614  & 0.0516 & 0.8148 \\
    IV & 27    & 133   & 0.0561 & 0.8223 \\
    V & 553   & 2,768  & 0.0447 & 0.8026 \\
    \bottomrule
    \end{tabular}%
  \label{t:simu_delta}%
\end{table}%

\begin{table}[htbp]
  \centering
  \caption{Sample Size \& Performance Using Standard Method}
    \begin{tabular}{lcccc}
    \toprule
    Case  & $n$   & Scenario & Type I error  & Power \\
    \midrule
    \multirow{2}[1]{*}{I} & \multirow{2}[1]{*}{1,472} & (i)   & \textcolor[rgb]{ 1,  0,  0}{0.1143} & \textcolor[rgb]{ 1,  0,  0}{0.758} \\
          &       & (ii)  & 0.0514 & \textcolor[rgb]{ 1,  0,  0}{0.681} \\
    \midrule
    \multirow{2}[1]{*}{II} & \multirow{2}[1]{*}{1,472} & (i)   & \textcolor[rgb]{ 1,  0,  0}{0.26} & \textcolor[rgb]{ 1,  0,  0}{0.697} \\
          &       & (ii)  & 0.0509 & \textcolor[rgb]{ 1,  0,  0}{0.5813} \\
    \midrule
    \multirow{2}[1]{*}{III} & \multirow{2}[1]{*}{6,040} & (i)   & \textcolor[rgb]{ 1,  0,  0}{0.0789} & \textcolor[rgb]{ 1,  0,  0}{0.7886} \\
          &       & (ii)  & 0.0494 & \textcolor[rgb]{ 1,  0,  0}{0.7323} \\
    \midrule
    \multirow{2}[1]{*}{IV} & \multirow{2}[1]{*}{95} & (i)   & \textcolor[rgb]{ 1,  0,  0}{0.1094} & \textcolor[rgb]{ 1,  0,  0}{0.7838} \\
          &       & (ii)  & 0.0585 & \textcolor[rgb]{ 1,  0,  0}{0.7156} \\
    \midrule
    \multirow{2}[2]{*}{V} & \multirow{2}[2]{*}{1,472} & (i)   & \textcolor[rgb]{ 1,  0,  0}{0.1504} & \textcolor[rgb]{ 1,  0,  0}{0.7301} \\
          &       & (ii)  & 0.0508 & \textcolor[rgb]{ 1,  0,  0}{0.6392} \\
    \bottomrule
    \end{tabular}%
  \label{t:simu_iid}%
\end{table}%

\section{Sample Size for Relative Lift}
\label{sec:relLiftSS}
\subsection{Motivation}
So far, sample size calculations are based on $\delta$ (i.e. $\mu_y - \mu_x$), often referred to as the ``absolute lift'' in A/B test. However, in practice the percentage treatment effect $\delta_{rel} = (\mu_y - \mu_x) / \mu_x$, often referred to as the ``relative lift'', is sometimes of more importance for both researchers and practitioners \cite{jiang2015asymptotic}.

We motivate this section with the following question -- when calculating sample size based on the relative lift $\delta_{rel}$, can we simply substitute $\delta = \delta_{rel} \cdot \mu_x$ into \eqref{iidSS}, \eqref{iidSS_bin}, and \eqref{deltaSS}? As the rest of this section will illustrate, although such an exercise is common in practice, it is not necessarily accurate in certain cases.

\subsection{Relative Lift from Independent Data}
\label{subsec:relSS_iid}
Let $\bar X = \sum_{i=1}^n X_i / n$ and $\bar Y =  \sum_{i=1}^n Y_i / n$ be the averages of the control and treatment observations, respectively, and $\hat \delta = \bar Y - \bar X$ and $\hat \delta_{rel} = (\bar Y - \bar X) / \bar X$ be the estimated absolute and relative lifts, respectively. 

For i.i.d. data, to calculate sample size based on the relative lift, we can first re-arrange \eqref{iidSS}, but in a slightly different way, as $\delta^2 / (2 \sigma^2/n) =  (z_{1-\alpha/2} + z_{1-\beta})^2$, which essentially equals:
\begin{equation}
\label{SSkey}
\left[\frac{\delta}{\se ( \hat \delta ) }\right]^2 =  (z_{1 - \alpha / 2} + z_{1 - \beta})^2.  
\end{equation}
With fixed type I error and power in the right hand side of \eqref{SSkey}, the left side entails $\delta$ from the alternative hypothesis and standard error of the observed metric. Similarly, in the case of relative lift, the corresponding formula becomes \footnote{The strict proof of \eqref{essense} is straightforward, and similar to the proof of standard sample size calculation \eqref{iidSS} that can be found in \cite{van2002statistical, kohavi2022b}, and hence omitted here.} :
\begin{equation}
    \left[\frac{\delta_{rel}}{\se(\hat{\delta}_{rel})}\right]^2 =  (z_{1-\alpha/2} + z_{1-\beta})^2,
    \label{essense}
\end{equation}
where, again, by the bi-variate Delta method \cite{deng2018}, which is similar to the one used in Section \ref{sec:deltaAM}, $\se(\hat{\delta}_{rel})$ can be estimated by:
\begin{equation}
     \se\left(\frac{\bar{Y}-\bar{X}}{\bar{X}} \right) = \se\left(\frac{\bar{Y}}{\bar{X}} \right) \approx 
    \left[ \frac{1}{n\mu_x^2} \left(\sigma_y^2 + \frac{\mu_y^2}{\mu_x^2} \sigma^2_x \right) \right]^{1/2}. 
    \label{relSE}
\end{equation}
The last step holds because treatment and control are independent, and therefore $\sigma_{xy}= 0$. 

We are now ready to calculate sample size for the relative lift. First, for continuous outcomes, under the same reasonable assumption of $\sigma_{y}^2 = \sigma_{x}^2 \equiv \sigma^2$ as used in the standard sizing formula \eqref{iidSS}, and from \eqref{essense} and \eqref{relSE} to solve $n_{rel}$, the required number of observations needed per arm to detect $\delta_{rel}:$
\begin{equation}
    n_{rel} = \left(\frac{1}{\mu_x^2} + \frac{\mu_y^2}{\mu_x^4}\right) \cdot \sigma^2 \cdot (z_{1-\alpha/2} + z_{1-\beta})^2 / \delta_{rel}^2.
    \label{relSS}
\end{equation}
Second, for binary outcomes, simply replacing $\mu_x$ and $\mu_y$ with $p_x$ and $p_y$ respectively, and replacing $\sigma^2$ with $p_\mathrm{pool}(1-p_\mathrm{pool})$, yields:
\begin{equation}
    n_{rel} = \left(\frac{1}{p_x^2} + \frac{p_y^2}{p_x^4}\right) \cdot p_\mathrm{pool}(1-p_\mathrm{pool}) \cdot (z_{1-\alpha/2} + z_{1-\beta})^2 / \delta_{rel}^2.
    \label{relSS_bin}
\end{equation}

\subsection{Comparison with Absolute Lift}
We now answer the motivating question at the beginning of this section. Because
$\delta_{rel} = \delta / \mu_x$, we can re-write \eqref{relSS} as
\begin{equation*}
n_{rel} = \left(1 + \frac{\mu_y^2}{\mu_x^2}\right) \cdot \sigma^2 \cdot (z_{1 - \alpha / 2} + z_{1 - \beta})^2 / \delta^2,
\end{equation*}
which differs from \eqref{iidSS} only in the first term: $1 + \mu_y^2 / \mu_x^2$ vs. $2$. Furthermore, because $\mu_y / \mu_x = \delta_{rel} +1$, the sample size difference is solely determined by $\delta_{rel}$ regardless of $\mu_x$. 

 For example, if a 1\% relative increase is anticipated in $H_1$, the sample size from (\ref{relSS}) is only 1\% higher than that from (\ref{iidSS}), which can be neglected. But if a 10\% relative increase is anticipated, $n_{rel}$ becomes 10.5\% higher than $n_{abs}$. If a 20\% relative increase is anticipated, $n_{rel}$ becomes 22\% higher than $n_{abs}$. These indicate that, if simply using the absolute lift sample size formula (\ref{iidSS}), the higher the relative lift in $H_1$, the more sample size will be under-estimated and hence the more under-powered the experiment will be. Also note that one may think a big percent change is rarely to be tested. However, the anticipated percent change, given it depends on the magnitude in the control arm, could be fairly large. For example, 20\% relative lift on a 5\% in control would mean only 1\% absolute lift. 

On the other hand, if $H_1$ is designed to detect negative lift as improvement, such as churn rate, the new relative sample size formula can be advantageous. For example, if a 1\% decrease is anticipated in $H_1$, the sample size from (\ref{relSS}) is 1\% less than that from (\ref{iidSS}), which is also negligible. But if a 10\% decrease is anticipated, $n_{rel}$ requires 9.5\% less than $n_{abs}$. If a 20\% decrease is anticipated, $n_{rel}$ needs 18\% less than $n_{abs}$.

The above phenomenon applies to both continuous and binary outcomes. The main reason for such a difference in sample size is that when relative lift is of interest, absolute difference based formula treats the mean value for control in the denominator as a constant instead of as a random variable, and the relative difference based formula derived as in (\ref{relSS}) does the latter. Since the true mean in control is always unknown, treating it as a random variable taking its uncertainty into account will give a more reliable estimate on the required sample size.

\subsection{Relative Lift from Correlated Data}
For clustered randomized experiments in Section \ref{sec:deltaSS}, we can estimate number of users instead by replacing $\sigma^2$ in \eqref{relSS} with $h$ as defined in \eqref{h}, and obtain:
\begin{equation}
\label{relSS_corrD}
k_{rel} = \left(\frac{1}{\mu_x^2} + \frac{\mu_y^2}{\mu_x^4}\right) \cdot h \cdot (z_{1 - \alpha / 2} + z_{1 - \beta})^2 / \delta_{rel}^2 ,
\end{equation} 
to detect relative lift. For binary outcomes, we can simply replace the general formula (\ref{relSS_corrD}) with
$\mu_x = p_x$, $\mu_y = p_y$ resulting in:
\begin{equation}
    k_{rel} = \left(\frac{1}{p_x^2} + \frac{p_y^2}{p_x^4} \right) \cdot h \cdot (z_{1-\alpha/2} + z_{1-\beta})^2 / \delta_{rel}^2.
    \label{relSS_corrD_bin}
\end{equation}

\subsection{Illustrating Example}
First, consider a case of independent data when the unit of randomization and the unit of analysis are both users. In order to test a relative lift of 10\%, if the current production of user-based $p_x = 0.6$, it translates to $p_y = (\delta_{rel} +1) p_x = 0.66$, and $p_\mathrm{pool} = (p_x + p_y) / 2 = 0.63$. Thus, from \eqref{relSS_bin}, $n_{rel} \approx 1,124$ users are needed for each arm.

Second, consider a case of correlated data when the randomization unit is user, whereas the analysis unit is session, using the same toy example data in Table \ref{t:deltaSSeg}, the current production session-based $p_x = \sum_i S_i / \sum_i N_i = 0.619$. Assuming 5\% relative lift in $H_1$ corresponds to $p_y = 0.65$, and from Section \ref{sec:deltaSSeg}, $h=0.151$. Thus, from (\ref{relSS_corrD_bin}), $k_{rel} \approx 2,603$ users are required for each arm.

\subsection{Simulation Studies}
\label{sec:sec4_simulation}
This section will focus on the i.i.d. case\footnote{Given the correlated data performance has been covered in the previous simulation section, the extension of relative lift for the cluster-randomized experiments becomes straightforward and hence is omitted here.} to illustrate the performance difference between absolute lift and relative lift. We still use binary outcome as the motivating example and start with true $p_x = 0.6$. In order to detect $\delta_{rel} = 10\%$ with 80\% power and 5\% type I error rate, using the proposed sample size method \eqref{relSS_bin}, we need $n_{rel} = 1,124$ per arm. We then simulate control arm from Binomial($n_{rel}$, $p_x$) and treatment arm from Binomial($n_{rel}$, $p_y$) respectively, constituting one data set. We assess the performance of type I error and power by independently generating 10,000 such data sets under the null and the alternative respectively. The true treatment relative lift is 0 under the null with $p_x = p_y = 0.6$ for both arms, and is $10\%$ under the alternative with $p_x = 0.6$, and $p_y = 0.66$. With 10,000 simulated data sets, the $80\%$ power and $5\%$ type I error estimates are subject to $\pm 0.8\%$ and $\pm 0.4\%$ margin of simulation error respectively. 

For each generated data set, we estimate the relative lift by $\hat{\delta}_{rel},$ and its corresponding standard error $\hat{\se}(\hat{\delta}_{rel})$ using \eqref{relSE} in Section \ref{subsec:relSS_iid}. Because the true mean and variance in \eqref{relSE} are typically unknown in practice when analyzing A/B data, the sample mean and variance are used as estimates. We deem the estimate statistically significant if  $|\hat{\delta}_{rel} / \hat{\se}(\hat{\delta}_{rel})| > z_{1-\alpha/2}.$ Aggregating across 10,000 data sets under $H_0$ and $H_1$ gives satisfactory type I error of 5.1\% and power of 81.3\%, respectively. In comparison, the absolute lift based sample size formula \eqref{iidSS_bin} would suggest that $1,017$ observations are required (i.e. $10.5\%$ less than the proposed method), consequently reducing the empirical power to 77.1\%. 

We extended this simulation example to a broad range of scenarios varying both true $p_x$ and $\delta_{rel}.$ For each scenario, we repeat the evaluation procedure outlined above, and report results in 
Tables \ref{t:sec4_simu_rel} and \ref{t:sec4_simu_abs}. There are two main take-aways:
\begin{itemize}
    \item First, Table \ref{t:sec4_simu_rel} shows that the proposed sample size method guarantees power and type I error rate across the board, whereas some cases in Table \ref{t:sec4_simu_abs} are under-powered, due to insufficient sample sizes estimated from the standard absolute lift based sample size;
    \item Second, it is confirmed that the impact is neglectable when the relative lift in $H_1$ is less than or equal to $5\%.$ However, the higher the $\delta_{rel}$ is, the more sample size is under-estimated by \eqref{iidSS_bin}, and the more under-powered the experiment is. 
\end{itemize}

\begin{table}[htbp]
  \centering
  \caption{Sample Size \& Performance Using Proposed Relative Lift Based Sample Size Method}
  \resizebox{0.43\textwidth}{!}{
    \begin{tabular}{lllll}
    \toprule
    True $p_x$ & True $\delta_{rel}$ in $H_1$ & n   & Type I error & Power \\
    \midrule
    \multirow{4}[2]{*}{0.1} & 1\%   &           1,433,336  & 0.0479 & 0.8019 \\
          & 5\%   &                 60,725  & 0.0512 & 0.8133 \\
          & 10\%  &                 16,301  & 0.0504 & 0.8210 \\
          & 20\%  &                   4,688  & 0.0529 & 0.8461 \\
    \midrule
    \multirow{4}[2]{*}{0.6} & 1\%   &              105,436  & 0.0525 & 0.7944 \\
          & 5\%   &                   4,342  & 0.0524 & 0.8064 \\
          & 10\%  &                   1,124  & 0.0505 & 0.8132 \\
          & 20\%  &                      299  & 0.0506 & 0.8399 \\
    \bottomrule
    \end{tabular}%
  \label{t:sec4_simu_rel}%
  }
\end{table}%

\begin{table}[htbp]
  \centering
  \caption{Sample Size \& Performance Using Standard Absolute Lift Based Sample Size Method}
  \resizebox{0.43\textwidth}{!}{
    \begin{tabular}{lllll}
    \toprule
    True $p_x$ & True $\delta_{rel}$ in $H_1$ & n     & Type I error & Power \\
    \midrule
    \multirow{4}[2]{*}{0.1} & 1\%   &           1,419,074  & 0.0530 & 0.8056 \\
          & 5\%   &                 57,764  & 0.0486 & 0.7949 \\
          & 10\%  &                 14,752  & 0.0507 & \textcolor[rgb]{ 1,  0,  0}{0.7838} \\
          & 20\%  &                   3,843  & 0.0562 & \textcolor[rgb]{ 1,  0,  0}{0.7599} \\
    \midrule
    \multirow{4}[2]{*}{0.6} & 1\%   &              104,387  & 0.0489 & 0.7946 \\
          & 5\%   &                   4,130  & 0.0526 & \textcolor[rgb]{ 1,  0,  0}{0.7875} \\
          & 10\%  &                   1,017  & 0.0514 & \textcolor[rgb]{ 1,  0,  0}{0.7714} \\
          & 20\%  &                      245  & 0.0540 & \textcolor[rgb]{ 1,  0,  0}{0.7565} \\
    \bottomrule
    \end{tabular}%
  \label{t:sec4_simu_abs}%
  }
\end{table}%

It is also noticed that the proposed method can over-power when $\delta_{rel}=20\%.$ 
Further investigation reveals that this is mostly due to use of sample (instead of true) mean in control when estimating variance of relative lift from \eqref{relSE} in A/B analysis. This can also explain why power is closer to desired $80\%$ when $\delta_{rel}$ is smaller. That is, the required sample size to detect a small relative lift is usually large enough to obtain a fairly accurate sample mean, and hence variance of the relative lift can be estimated reasonably accurately, thus the power is closer to desired $80\%$. However, when $\delta_{rel}$ to be detected is large, like $20\%$, the required sample size becomes much smaller and the sample mean estimate is no longer as accurate. To verify our intuition, we replace only the sample mean in control with the true mean in the variance estimation and rerun the simulation analyses. The results shown in Table \ref{t:sec4_simu_rel2} before each bracket have the power approaching more closely to the desired $80\%$ level. To mimic reality, instead of using the true mean, we also tentatively drew 5000 samples as if it is collected from the historical data and used its sample mean for $\mu_x$ in the variance estimation, and it has almost the same performance shown in brackets in Table \ref{t:sec4_simu_rel2}. In all, if the relative lift to be detected is large and an exact designed power is desired, a large sample mean from the historical data is recommended in variance estimation in A/B analysis. Otherwise, a regular variance estimation process using sample means from the experiment's data still warrants at least $80\%$ power. Both cases outperform the standard absolute lift based sample size method. 
\begin{table}[htbp]
  \centering
  \caption{Performance under Proposed Method \\- using true $\mu_x$ or large-sample mean in variance estimation}
  \resizebox{0.48\textwidth}{!}{
    \begin{tabular}{llll}
    \toprule
    True $p_x$ & True $\delta_{rel}$ in $H_1$ & \multicolumn{1}{p{8.165em}}{Type I error \newline{}true (large-sample)} & \multicolumn{1}{p{8.165em}}{Power\newline{}true (large-sample)} \\
    \midrule
    \multirow{4}[2]{*}{0.1} & 1\%   & 0.048 (0.052) & 0.8 (0.799) \\
          & 5\%   & 0.05 (0.051) & 0.808 (0.805) \\
          & 10\%  & 0.052 (0.052) & 0.81 (0.803) \\
          & 20\%  & 0.056 (0.057) & 0.819 (0.815) \\
    \midrule
    \multirow{4}[2]{*}{0.6} & 1\%   & 0.052 (0.053) & 0.793 (0.792) \\
          & 5\%   & 0.053 (0.052) & 0.798 (0.8) \\
          & 10\%  & 0.052 (0.054) & 0.802 (0.802) \\
          & 20\%  & 0.054 (0.054) & 0.812 (0.812) \\
    \bottomrule
    \end{tabular}%
  \label{t:sec4_simu_rel2}%
  }
\end{table}%

\section{What minimum observed difference will be statistically significant?}
\label{sec:MOD}
This question is often asked by stakeholders and the answer is not the average treatment effect (ATE, a.k.a. MDE) used to power the experiment, but can be answered at the design stage.
Moreover, when it comes to the sample size calculation, for the situation of the required ATE not available,  is there an alternative way to compute the sample size? We will answer these two questions in this section.

\subsection{Difference between ATE and MOD}
We will use $\Delta_\mathrm{ATE}$ and $\Delta_\mathrm{MOD}$ to denote the lift for the two different concepts respectively in this section. $\Delta_\mathrm{ATE}$, which is the same as $\delta$ or $\Delta_\mathrm{MDE}$, is predominately used at the design stage to power the experiment. It is the best guess of the magnitude of the lift. We have reviewed and extended sample size formulas \eqref{iidSS}, \eqref{iidSS_bin} and \eqref{deltaSS}, to ensure there is a high probability (e.g., 80\%) that the experiment will show statistical significance, if the true underlying treatment effect is larger than or equal to ATE. For simplicity, we will focus on absolute lift hereafter.

$\Delta_\mathrm{MOD}$, however, is the minimum observed difference that will be statistically significant. This is the smallest observed treatment effect that would be expected to result in a $p< 0.05$\footnote{This concept is from the pharmaceutical industry, and to our best knowledge, not widely discussed or adopted within the A/B testing community.}. It is important to ensure that $\Delta_\mathrm{MOD}$ is large enough to be business meaningful, as we do not want an experiment to show a statistically significant result where the observed difference will not impact business much.

\subsection{MOD Determination}
$\Delta_\mathrm{MOD}$ can actually be estimated at the design stage, given $\Delta_\mathrm{ATE}$, desired power and type I error rate:
\begin{equation}
|\Delta_\mathrm{MOD}| \approx \frac{z_{1 - \alpha / 2}}{z_{1-\alpha / 2} + z_{1 - \beta}} |\Delta_\mathrm{ATE}|.
\label{MODvsATE}
\end{equation}

To prove \eqref{MODvsATE}, on one hand, $\Delta_\mathrm{ATE}$ by definition corresponds to the sample size calculation in \eqref{SSkey}, namely:
\begin{equation}
\left|\frac{\Delta_\mathrm{ATE}}{\se(\hat \delta)}\right| =  z_{1 - \alpha / 2} + z_{1-\beta}.
\label{SS_ATE}
\end{equation}
On the other hand, at the end of the experiment, in order to be statistically significant, $\Delta_\mathrm{MOD}$ by definition lies at the boundary of the null hypothesis rejection region, namely:
\begin{equation}
\left|\frac{\Delta_\mathrm{MOD}}{\se(\hat \delta)}\right| =  z_{1 - \alpha / 2}. 
\label{SS_MOD}
\end{equation}
Notice that $\Delta_\mathrm{MOD}$ does not depend on power. Combining \eqref{SS_ATE} and \eqref{SS_MOD} results in \eqref{MODvsATE}. 

With 80\% power and 2-sided 5\% type I error rate, \eqref{MODvsATE} gives $|\Delta_\mathrm{MOD}| \approx 0.7 |\Delta_\mathrm{ATE}|$. When power is set to be 90\% with type I error unchanged, the ratio becomes 0.6. This MOD question is often asked in the context of assuming $\Delta_\mathrm{MOD}$ has the same sign direction as $\Delta_\mathrm{ATE}$, and not so much about the significant opposite impact, so the absolute sign can be dropped for simplicity as a rule of thumb.

Besides providing the solution to the title question of this section, for the sample size calculation, when $\Delta_\mathrm{ATE}$ is not available, but $\Delta_\mathrm{MOD}$ is known under business context, \eqref{MODvsATE} can also be used to back calculate $\Delta_\mathrm{ATE}$ as a rough estimate, which can then be used to compute the sample size.

\subsection{Illustrating Example}
For continuous outcomes, take an example of a user engagement metric, such as average time played, at the design stage, the anticipated mean change from the control to the treatment group are 3 to 3.5 units. 
For this design, the estimated $\Delta_\mathrm{MOD} \approx 70\% \cdot (3.5-3) = 0.35$. That is, if the experiment were powered at 80\% with a 2-sided p=0.05 and the anticipated treatment benefit was $\Delta_\mathrm{ATE} = 0.5$, we would estimate at the design stage that any difference of 0.35 or greater in the means would yield an observed $p<0.05$.

For binary outcomes, assume the expected conversion rate at the design stage for the control group and the treatment group are 10\% and 12\% respectively. So $\Delta_\mathrm{ATE} = 2\%$ is the absolute difference.  If the experiment is sized to have 80\% power at a 2-sided $p=0.05$ level,  the estimated $\Delta_\mathrm{MOD} \approx 70\% \cdot 2\% = 1.4\%$. This approximation works well for binomial data, with an exception being when the actual rates are grossly under- or over-estimated.

\subsection{Simulation Studies}
\label{sec:sec5_simulation}
\begin{figure}[h]
  \centering
  \includegraphics[width=8.58cm]{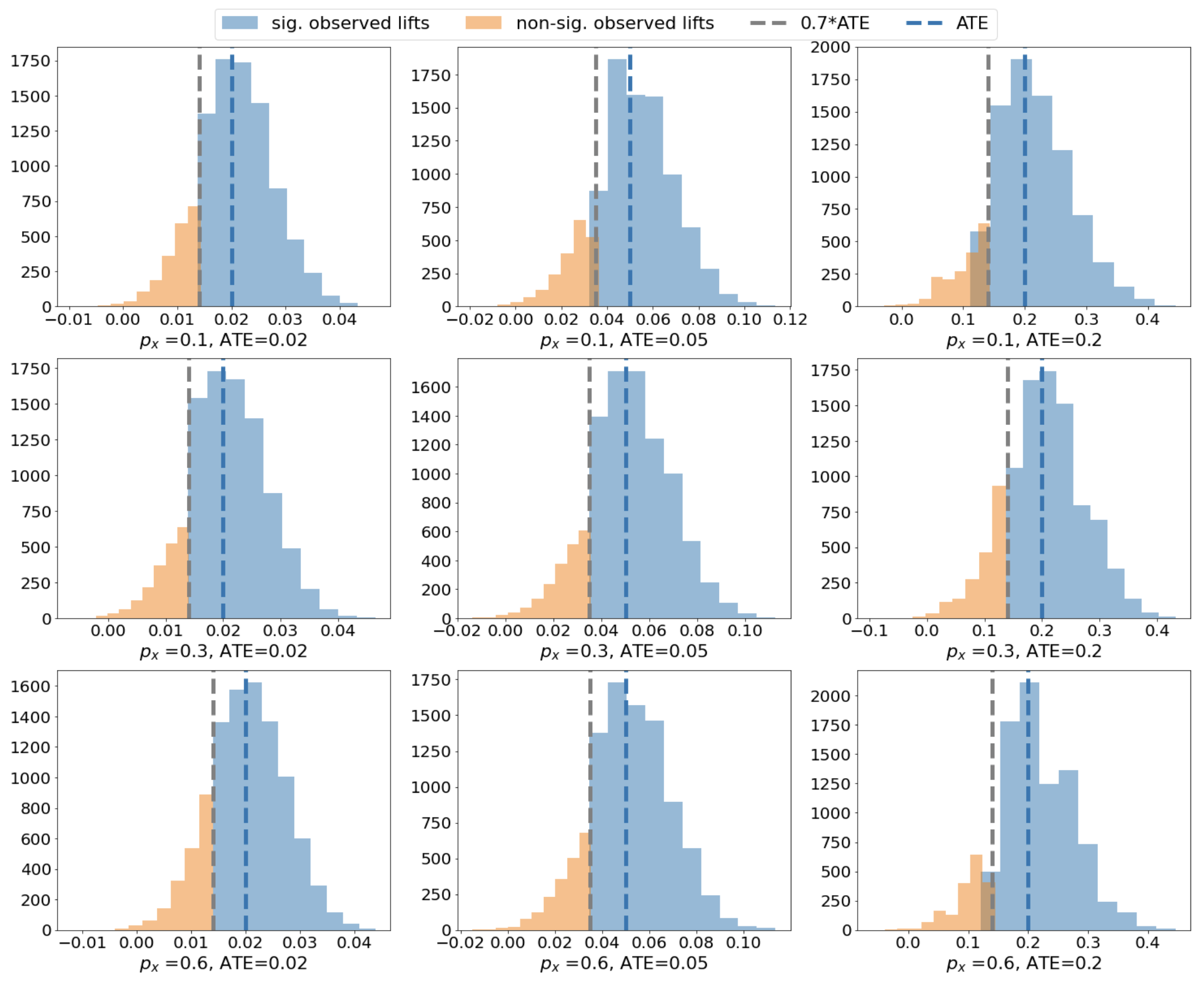}
  \caption{MOD vs ATE: significant and non-significant observed lifts cross 10,000 simulations}
  \Description{MOD vs ATE}
  \label{f:sec5_simu}
\end{figure}

We now conduct a simulation study to verify the MOD and ATE relationship as claimed. Again, we will focus on the binary outcome case. For simplicity, we will demonstrate the performance under the absolute lift and independent data scenario. Assuming a true mean for control $p_x$ and a desired ATE to be detected, we first compute the required sample size $n$ using the standard sample size \eqref{iidSS_bin}. 10,000 simulation replicates are considered, with data in each replicate consisting of Binomial($n$, $p_x$) for control and Binomial($n$, $p_y$) for treatment, where $p_y$ can be inferred from ATE and $p_x$. We then identify all the observed lifts across all the simulation replicates when the statistical significance is detected through the t-test statistic for the observed lift crossing the critical value, i.e. $|\hat{\delta} / \hat{\mathrm{SE}}(\hat{\delta})| > Z_{1-\alpha/2}$. Likewise, all the non-significant observed lifts are recorded as well. We plotted all 10,000 observed lifts as a histogram with significant ones shown in blue and non-significant ones in orange for each pair of \{true $p_x$, ATE\}. This was done for a total 9 different pair scenarios, all of which are shown in Figure \ref{f:sec5_simu}. Additionally, a vertical blue line is placed in each plot to represent ATE used in the power analysis, and $0.7 *$ ATE is displayed as the grey line. 

From the 9 scenarios, the $0.7*$ATE line well separates the significant observed lifts from the non-significant ones, which exactly answers the titled question of this section. There are cases in which a small amount of significant observed lifts overlap with non-significant ones. This is expected due to the statistical significance not only relying on the magnitude of the observed lift, but also depending on its variance. 
In those cases, $0.7*$ ATE line robustly picks the more conservative minimal value such that it is unlikely to see the same observed lift to be non-significant in a different experiment. All the above findings successfully verify the validity of MOD concept and its relationship to ATE.

\section{Best Practices}
\subsection{Is Balanced Design Always the Best?}
\label{sec:Nf}
As noted in Section \ref{sec:iid}, with the intent of minimizing overall sample size, a one-to-one sample size ratio for the treatment and control groups is often assumed before computing the required sample size per arm. However, in online controlled experiments, samples in control are usually cheap to collect without much additional operational cost and do not hurt users' experience. Thus, the corresponding required A/B experiment duration actually depends on the required sample size in the treatment group, and not the total sample size for both arms. Therefore, minimizing the sample size in treatment is the key to shortening the experiment duration, if duration becomes a major concern from a regular balanced design. To achieve this, the sample size Equation (\ref{SSkey}) is needed and can be generalized to a form without the balanced design assumption: 
$\frac{\delta^2}{(\frac{1}{n_x} +\frac{1}{n_y})\sigma^2 } =  (z_{1-\alpha/2} + z_{1-\beta})^2$,     
which can be rewritten as:
\begin{equation}
     \mathrm{Power} = 1- \Phi^{-1}(z_{1-\alpha/2} - \delta\sqrt{n_{all}\cdot f(1-f)}/\sigma),
    \label{powerNf}   
\end{equation}
where we let $n_{all} = n_x + n_y$ denote the total sample size for both arms,  $f=n_y / n_{all}$ denote the sample allocation proportion for the treatment group, and $\Phi^{-1}(\cdot)$ is inverse cumulative density function for standard normal. It can be seen that power is essentially a function of $n_{all}\cdot f(1-f)$, when other design parameters $\alpha$, $\delta$ and $\sigma$ are fixed. It is immediate that when $n_{all}$ is also fixed, power is maximized when $f=0.5$, which explains why a balanced design is typically recommended. When there is more flexibility to enlarge sample size in control, however, this implies $n_{all}$ is no longer fixed. Consequently, the minimizing duration aim distills downs to minimizing sample size in treatment (i.e. $n_{all}\cdot f$) while keeping power Equation (\ref{powerNf}) at a fixed level, say 80\%. Solving this optimization problem leads to the conclusion that the smaller the $f$ is, the shorter the duration will be. This can be achieved by shrinking sample size in treatment, but with the cost of total sample size being larger than before. That is, the enlargement in control sample size will be larger than the shrinkage in treatment sample size. This may be still worth it, for example, if we allow $n_x = 2n_y$ (i.e. $f=1/3$) allocated each day, it can shorten the duration by 25\% from a balanced design, but the total sample size will be 12.5\% larger than original balanced design; or setting $n_x = 4n_y$ (i.e. $f=0.2$) will shorten the duration by 37.5\%, but the total sample size will be 56.25\% larger than original total sample size. In all, under the context of non-costly control allocation, these duration gains will help test any new feature faster and benefit both users and companies with faster launching. One caveat is that this optimization is under a typical common assumption of equal variance of the treatment and control groups. While this assumption is often true, or nearly so in practice, we should still keep a decent amount of traffic in the treatment group in order to have reliable point estimate and variance of the metric in the treatment group. Hence, a traffic allocation proportion $f<0.2$ is not recommended. This recommendation also comes from the fact that the duration gain becomes less and less efficient as traffic allocation proportion is reduced. For example, when a more extreme $n_x = 9n_y$ (i.e. $f=0.1$) is set, even though the duration can be cut down 44.4\% from the balanced design, the total sample size will be 178\% larger than the balanced design.

\subsection{Does Skewed Data matter?}
The normal assumption is well-known for a regular two-sample t-test, and hence it is often seen that experimenters will instead choose a non-parametric test (e.g. Wilcoxon rank-sum) \cite{hettmansperger2010} and corresponding non-parametric sample size estimation \cite{divine2010} when they see the data are not normally distributed. However, is this a necessary practice? There is a common confusion as to what base this normal assumption refers to. In fact, the normal assumption is with respect to the metric to be tested such as a mean or a rate, not necessarily on the data itself. According to central limit theorem \cite{van2000clt}, a sample of 30 or more can generally be considered large enough to ensure the mean (including rate) metric follows a normal distribution. To gain more confidence, the metric can also be proven roughly normal by bootstrapping the metric thousands of times from the same skewed data, and then assessing the metric distribution. With that, a regular t-test can be still used without concerns and hence a non-parametric approach may not be needed. In the merit of an online experiment, having only a small sample size available to test is rare, and hence, a regular t-test will often be sufficient, even when the data are observed skewed. 

It is noted that it is not incorrect to use a non-parametric test and corresponding sample size estimation, but it is known to be less powerful than a regular t-test, and it is testing the equality of the median or the whole distribution between treatment and control, rather than the mean of a continuous or binary outcome. However, the skewed data are not without drawbacks. The main one is the variance of the metric becomes a lot larger when the data are skewed, which will result in a larger required sample to detect the same lift of interest. When the required sample or duration becomes challenging to implement, it can be mitigated in different ways, such as using capping, binarization, using conversion rate metrics instead of mean spent metrics, etc. \cite{kohavi2020trustworthy}.

\subsection{Total Metrics vs Mean Metrics}
From the perspective of business partners, the total metrics such as total downloads, total sales, total billing are often easier to understand when comparing the treatment and the control group to assess how well the new feature performed in a A/B testing. But, is it legitimate for such "total metrics" to be tested and sized? There are two main problems. 
First, the two-sample t-test was developed to test means of two populations, not sums of two populations or two samples. When relative lift is of interest, and the sample sizes are exactly the same for the treatment and control groups, testing on the mean is the same as testing on the sum because the denominator can be cancelled in both arms. However, in online controlled experiments, it is difficult to control the exact number of samples allocated for each arm. From practical experience, it is often seen that the two arms are roughly similar but never identical. Sometimes only impressed or triggered samples are logged and tracked, and hence, it is not possible to guarantee the allocated samples between two arms are identical. In those cases, it is dangerous to directly compare the non-normalized metric, the total, since the actual unequal sample size for each arm plays a role in the total metric and hence may impact the significance or even drive the conclusion in the opposite direction. 

The second problem is related to the estimation of its variance when absolute lift is of interest. For each arm, the variance of the total metric, under the simplest i.i.d. case, $\var(\sum_{i=1}^{n} X_i) = n\sigma^2$ will increase as sample size increases, whereas for a mean metric $\var(\bar X) = \sigma^2/ n$ will decrease as sample size grows. The same conclusion applies to correlated data as well. This illustrates another reason total metrics should be discouraged. 

In summary, mean metrics for any type of outcome do not carry any of the above problems, and should be always recommended as the metric rather than total metrics.

\subsection{Other Sample Size Calculation Variants}
Another observed bad practice is to use the confidence interval of the lift directly to compute the required sample size. For example, if we want to test an absolute lift of 0.05 from the conversion rate in control at 0.3. With the intent of a significant result, assuming a balanced design and i.i.d. for simplicity, the lower bound of the confidence interval, $0.05-z_{1-\alpha/2}\sqrt{2\sigma^2/n}$, needs to be larger than 0, where $\sigma^2 = p_\mathrm{pool}(1-p_\mathrm{pool})$ and $p_\mathrm{pool} = (0.3+0.35)/2$. From this, the sample size per arm, $n$, can be computed. The problem of this approach is that there is no power involved. Additionally, the corresponding estimated sample size will actually be related to the MOD concept in Section \ref{sec:MOD}, and smaller than the sample size required from Equation (\ref{iidSS}) that will ensure desired power while controlling type I error. One may also try to infer the confidence interval width, or the marginal error for the lift from a previous experiment and leverage it for the new experiment design. The underlying problem with this is that the $\sigma^2$ cannot be directly borrowed as it depends on the baseline control rate as well as the observed treatment lift in the previous experiment, both of which can be changed for the new experiment even for the same type of metric. Therefore, using the standard path of sample size calculation shown in Section \ref{sec:exist}-\ref{sec:relLiftSS} is highly recommended.

\section{Concluding Remarks}
\label{sec:conclusion}

In the era of big data, A/B testing has become an increasingly integral and mainstream tool for modern software development. There has been existing literature focusing on analyzing large-scale experiments, but an important counterpart of robust online experimental design, sample size calculation, is often neglected. To fill this gap, this paper proposes novel extensions on sample size calculations to specifically address correlations within randomization units, and to properly assess relative vs. absolute lift. Moreover, we demystified a common phenomenon in practical A/B testing, where the estimated treatment effect, while less than pre-specified average treatment effect, can be nevertheless statistically significant. We explained the underlying reasons hoping to provide insights on future experiment designs. Lastly, we discussed common pitfalls, and provided strategies for trustworthy sample size calculations and experimental designs. Although we focus on the fixed-horizon A/B testing in this paper, we believe that such techniques are readily applicable to more advanced designs, such as sequential \cite{mehta2001, maharaj2023anytime}, or more complex scenarios with user behaviour \cite{wang2019heavy} or temporal \cite{sadeghi2022novelty} heterogeneity, or where percentiles \cite{wang2021conq} are of interest. We leave them as future directions.

\appendix
\section{Appendix}

\subsection{Proof of Theorem \ref{thm:equiv}}

We first re-write the components in \eqref{deltaXbar} as:
\begin{equation}
\label{mus}
\mu_S 
= \e
\left[ 
\e 
\left( 
\sum\nolimits_j X_{ij} \mid \mu_i, \sigma_i, N_i 
\right) 
\right] 
= \e (N_i \mu_i),
\end{equation}
\begin{align}
\label{sigmas2}
\sigma^2_S 
& = 
\e \left[ \var \left( \sum\nolimits_j X_{ij} \mid \mu_i, \sigma_i, N_i  \right) \right]
+
\var \left[ \e \left( \sum\nolimits_j X_{ij} \mid \mu_i, \sigma_i, N_i  \right) \right] \nonumber \\
& = \e ( N_i \sigma_i^2 ) + \var ( N_i \mu_i ) \nonumber \\
& = \e ( N_i \sigma_i^2 ) + \e ( N_i^2 \mu_i^2 ) - \left[ \e (N_i \mu_i) \right]^2 \nonumber \\
& = \e ( N_i \sigma_i^2 ) + \e ( N_i^2 \mu_i^2 ) - \mu_S^2,
\end{align}
and
\begin{align}
\label{sigmasn}
\sigma_{SN} 
& = \e(S_i N_i) - \e(S_i)\e(N_i) \nonumber \\
& = \e \left[ \e \left( N_i \sum\nolimits_j X_{ij} \mid \mu_i, \sigma_i, N_i  \right) \right] - \mu_S \mu_N \nonumber \\
& = E( \mu_i N_i^2 ) - \mu_S \mu_N.
\end{align}

With \eqref{mus}--\eqref{sigmasn}, we are ready to prove Theorem \ref{thm:equiv}. First, for the control group, we estimate each term in \eqref{deltaXbar}, denoted as
$
\hat \mu_N, \hat \sigma_N, \hat \mu_S, \hat \sigma_S
$
and
$
\hat \sigma_{SN},
$
respectively. Second, assuming no effect on $N_i,$ the corresponding treatment mean and variance stay the same:
\begin{equation}
\label{updatemunandsigman}    
\hat \mu_N(T) = \hat \mu_N, 
\quad
\hat \sigma_N(T) = \hat \sigma_N.
\end{equation}
Third, from \eqref{mus}
\begin{equation}
\label{updatemus}
\hat \mu_S(T) = \hat \mu_S + \delta   \hat \mu_N,
\end{equation}
and from \eqref{sigmasn}
\begin{equation}
\label{updatesigmasn}
\hat \sigma_{SN}(T) = \hat \sigma_{SN} + \delta   \hat \sigma_N^2.
\end{equation}
Finally, assuming no effect on $\sigma_i$ and by \eqref{sigmas2},
\begin{align}
\label{sigmas2diff}
\sigma_S^2(T) - \sigma_S^2(C) 
& = \e \left[ N_i^2 \left( \mu_i + \delta \right)^2 \right] - \mu_S^2(T) - \e ( N_i^2 \mu_i^2 ) + \mu_S^2(C) \nonumber \\
& = 2\delta \cdot \e ( \mu_i N_i^2 ) + \delta^2   \e ( N_i^2 ) + \mu_S^2(C) - \mu_S^2(T) \nonumber \\
& = 
2\delta   \left[ \sigma_{SN}(C) + \mu_S(C)\mu_N(C) \right] \nonumber \\ 
& + 
\delta^2  \left[ \sigma_N^2(C) + \mu_N^2(C) \right]
+ 
\mu_S^2(C) - \mu_S^2(T).
\end{align}
The last steps holds because of \eqref{sigmasn}. Consequently,
\begin{align}
\label{updatesigmas2}
\hat \sigma_S^2(T) 
& = 
\hat \sigma_S^2
+
2\delta   \left[ \hat \sigma_{SN} + \hat \mu_S \hat \mu_N \right]
+ 
\delta^2  \left[ \hat \sigma_N^2 + \hat \mu_N^2 \right]
+ 
\hat \mu_S^2 
- 
\left(
\hat \mu_S + \delta   \hat \mu_N
\right)^2
\nonumber \\
& = 
\hat \sigma_S^2
+
2\delta \hat \sigma_{SN} 
+ 
\delta^2  \hat \sigma_N^2.
\end{align}
The last step holds because of \eqref{updatemus}.

Effectively, \eqref{updatemunandsigman}--\eqref{updatesigmas2} ``updated'' each term in \eqref{deltaXbar} for the treatment group. To be more specific, 
\begin{align}
\label{vartreatment}
\widehat \var(\bar Y) 
& = 
\frac{1}{k \hat \mu_N^2(T)} 
\left[
\hat \sigma_S^2(T) 
-2
\frac{
\hat \mu_S(T)
}{
\hat \mu_N(T)
}
\hat  \sigma_{SN} (T)
+
\frac{
\hat \mu_S^2  (T)
}
{
\hat \mu_N^2  (T)
}
\hat \sigma_N^2  (T)
\right] \nonumber \\
& = 
\frac{1}{k \hat \mu_N^2} 
\left[
(
\hat \sigma_S^2
+
2\delta \hat \sigma_{SN} 
+ 
\delta^2  \hat \sigma_N^2
)
-2
\frac{
\hat \mu_S + \delta   \hat \mu_N
}{
\hat \mu_N
}
(
\hat \sigma_{SN} + \delta   \hat \sigma_N^2
)
\right] \nonumber \\
& +
\frac{1}{k \hat \mu_N^2} 
\left[
\frac{
\left(
\hat \mu_S + \delta   \hat \mu_N
\right)^2
}
{
\hat \mu_N^2
}
\hat \sigma_N^2
\right] \nonumber \\
& = 
\frac{1}{k \hat \mu_N^2} 
\left[
\hat \sigma_S^2
-
\delta^2  \hat \sigma_N^2
-2
\frac{
\hat \mu_S
}{
\hat \mu_N
}
\hat \sigma_{SN}
-2 \delta
\frac{
\hat \mu_S
}{
\hat \mu_N
}
\hat \sigma_N^2
\right] \nonumber \\
& +
\frac{1}{k \hat \mu_N^2} 
\left[
\frac{
\hat \mu_S^2 + 2\delta \hat \mu_S  \hat \mu_N + \delta^2 \hat \mu_N^2
}
{
\hat \mu_N^2
}
\hat \sigma_N^2
\right] \nonumber \\
& = 
\frac{1}{k \hat \mu_N^2} 
\left[
\hat \sigma_S^2
-2
\frac{
\hat \mu_S
}{
\hat \mu_N
}
\hat \sigma_{SN}
+
\frac{
\hat \mu_S^2
}
{
\hat \mu_N^2
}
\hat \sigma_N^2
\right].
\end{align}
The proof is complete. 

\subsection{Proof of Theorem \ref{thm:bin}}

The proof is rather similar to that of Theorem \ref{thm:equiv}. First, \eqref{updatemunandsigman}-\eqref{updatesigmasn} still hold, because they are not at all related to assumption of no effect on $\sigma_i.$ Second, by \eqref{sigmas2} and because $\sigma_i^2 = p_i(1 - p_i),$ the difference between
$
\sigma_S^2(T)
$
and
$
\sigma_S^2(C),
$
besides \eqref{sigmas2diff}, has an additional term
\begin{align*}
\e \left[ N_i (p_i + \delta)(1 - p_i - \delta) \right] - \e \left[ N_i p_i (1 - p_i) \right]
&= \e \left[ N_i \delta (1 - \delta - 2 p_i) \right] \\
&= \delta(1 - \delta) \mu_N - 2\delta \mu_S.
\end{align*}
Consequently, \eqref{vartreatment} ``almost'' holds except that we need to account for the additional term. To be specific, 
\begin{align*}
\label{vartreatment-bin}
\widehat \var(\bar Y) 
& = 
\frac{1}{k \hat \mu_N^2} 
\left\{
\left(
\hat \sigma_S^2
-2
\frac{
\hat \mu_S
}{
\hat \mu_N
}
\hat \sigma_{SN}
+
\frac{
\hat \mu_S^2
}
{
\hat \mu_N^2
}
\hat \sigma_N^2
\right)
+ 
\delta
\left[
(1 - \delta) \hat \mu_N - 2 \hat \mu_S
\right]
\right\}.
\end{align*}
The proof is complete.

\clearpage
\bibliographystyle{ACM-Reference-Format}
\balance
\bibliography{sigproc}

\end{document}